\documentclass[twocolumn,secnumarabic,amssymb, nobibnotes, aps, prl]{revtex4-2}

\usepackage{graphicx}
\usepackage{dcolumn}
\usepackage{bm}
\usepackage[hidelinks]{hyperref}

\usepackage{subcaption}
\usepackage[english]{babel}
\usepackage{amsmath}
\usepackage{caption}
\begin{document}

\title{
Photon-induced $J/\psi$ as a linearly polarized probe of collision geometry\\
in relativistic heavy-ion collisions
}

\collaboration{STAR Collaboration}

\begin{abstract}

In relativistic heavy-ion collisions, the linear polarization of quasi-real photons can correlate the spin orientation of photon-induced vector mesons with the collision geometry.
To explore this sensitivity, we present the first measurement of the decay angular distribution of photon-induced \(J/\psi\) with respect to the event plane in heavy-ion collisions with hadronic overlap.
The analysis uses Ru+Ru and Zr+Zr collisions at a nucleon--nucleon center-of-mass energy of \(\sqrt{s_{\mathrm{NN}}}=200\) GeV recorded by STAR.
After correcting for the finite event-plane resolution and subtracting the residual hadronic contribution, we extract a negative second-order modulation,
\(A_2^{\rm phot} = -0.39 \pm 0.11~(\mathrm{stat.}) \pm 0.04~(\mathrm{sys.})\),
for \(|y^{ee}|<0.8\), \(p_{\mathrm T}^{ee}<0.2~\mathrm{GeV}/c\), and 30--80\% centrality.
This modulation provides evidence that the \(J/\psi\) polarization is correlated with the event plane, as expected for linearly polarized photons.
The observed modulation establishes polarized \(J/\psi\) photoproduction as a direct probe of the transverse orientation of the initial collision geometry.

\end{abstract}

\maketitle

Relativistic heavy-ion collisions not only create the hottest strongly
interacting matter~\cite{Braun-Munzinger:2007edi} in the laboratory but also generate the strongest electromagnetic fields~\cite{Skokov:2009qp,Deng:2012pc,Huang:2015oca,Hattori:2016emy,Shen:2025unr} known in nature. In the equivalent-photon approximation~\cite{Bertulani:2005ru}, the intense Lorentz-contracted Coulomb fields of the fast-moving nuclei can be treated as high-flux beams of quasi-real photons with negligible virtuality, giving rise to a broad class of photon-induced processes in nucleus--nucleus collisions.
These processes include two-photon interactions,
$\gamma\gamma$~\cite{Baur:2001jj,Klein:2020fmr}, and photonuclear reactions,
$\gamma A$~\cite{Klein:1999qj,Baur:2001jj,Klein:2020fmr}.
In coherent photonuclear production, the photon interacts with the nucleus as a whole, which remains intact~\cite{Klein:1999qj,Toll:2012mb}; in incoherent production, the photon interacts with an individual nucleon and generally leaves the nucleus excited or broken up~\cite{Good:1960ba,Klein:1999qj}. Photon-photon interactions provide clean access to strong-field QED~\cite{Baltz:2007gs,Klusek-Gawenda:2016nuo,Zha:2018tlq,Zha:2021jhf,Brandenburg:2021lnj,Li:2023yjt,Zha:2023sed}, while photonuclear reactions probe the gluonic structure of nuclei~\cite{Lappi:2010dd,Adeluyi:2012ph,Guzey:2013xba,Frankfurt:2015cwa,Wu:2025dxg}.

A more recently appreciated feature of photon-induced processes in heavy-ion collisions is the distinct linear polarization of the quasi-real photons.
Arising from the Lorentz-contracted Coulomb fields of the relativistic nuclei, this polarization is oriented in the transverse plane, with the electric-field vectors pointing radially outward from the source nucleus.
Such polarization imprints itself onto the final-state particles, resulting in characteristic angular modulations.
In the Breit--Wheeler process ($\gamma\gamma \to e^{+}e^{-}$), this polarization produces a distinctive fourth-order modulation in the lepton azimuthal distribution, measured in the pair rest frame with respect to the pair transverse momentum~\cite{Li:2019yzy,Li:2019sin,Xiao:2020ddm}.
Following an analogous definition in vector-meson photoproduction, a second-order modulation has been observed in the decay angular distribution.
The underlying physics, however, is richer: the vector meson can inherit the linear polarization of the incident photon, while quantum interference between the two indistinguishable production amplitudes from the two nuclei can correlate this polarization direction with the vector-meson transverse momentum~\cite{Klein:1999gv}.
This interplay results in a characteristic spin-interference pattern~\cite{Zha:2020cst,Xing:2020hwh,Brandenburg:2022jgr}.

These polarization-induced patterns have been confirmed experimentally, primarily in ultra-peripheral collisions (UPCs), where the impact parameter exceeds the sum of the nuclear radii and hadronic interactions are absent.
They have been observed by STAR in the Breit--Wheeler channel~\cite{STAR:2019wlg}, by STAR and ALICE in coherently photoproduced $\rho^{0}$ mesons~\cite{STAR:2022wfe,ALICE:2024aza}, and most recently by STAR for $J/\psi$ mesons~\cite{STAR:2025wpi}.
In these measurements, the polarization-sensitive angle is defined with respect to the transverse momentum of the reconstructed final-state pair, which serves as an experimentally accessible kinematic reference direction.
However, these measurements do not directly establish the expected connection between the photon-induced linear polarization and the initial collision geometry, characterized by the impact-parameter vector and the associated reaction plane.
This raises a central question: can this polarization direction be probed directly with respect to the collision geometry?

\begin{figure}[htb]
\centering
\includegraphics[width=\columnwidth]{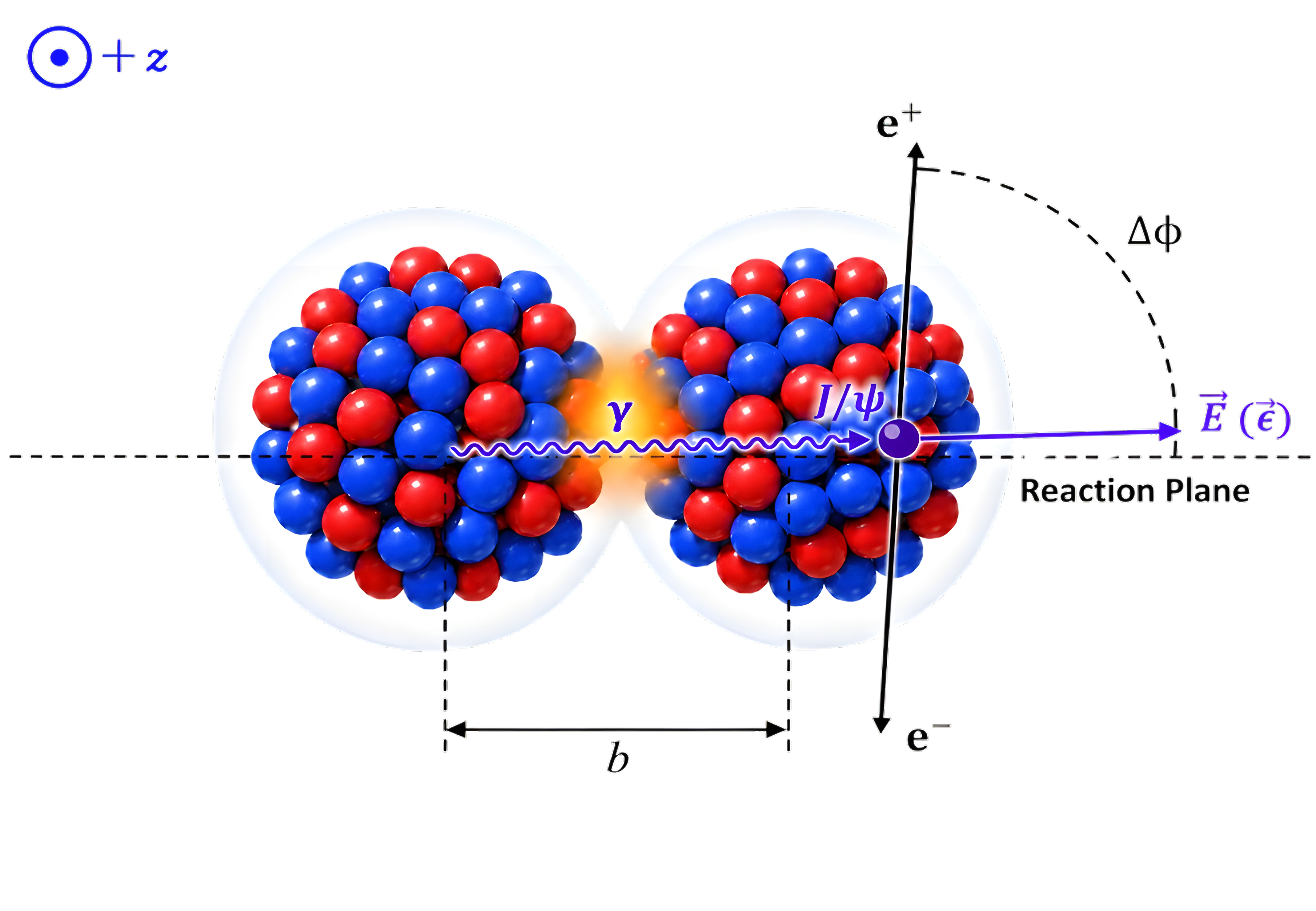}
\caption{
Schematic illustration of coherent photonuclear \( J/\psi \) production in the transverse plane.
The photon electric-field direction (\(\vec{E}\)), which specifies its linear-polarization vector (\(\vec{\epsilon}\)), is approximately aligned with the impact-parameter vector (\(\vec{b}\)).
The produced \(J/\psi\) inherits this linear polarization, and \(\Delta\phi\) denotes the azimuthal angle of a decay lepton relative to the reaction plane.
}

\label{fig:diagram}
\end{figure}

A proposed route to such a direct measurement is to study photonuclear $J/\psi$ production relative to a reaction-plane reference~\cite{Wu:2022exl}. As illustrated in Fig.~\ref{fig:diagram}, this polarization axis is approximately aligned with the impact-parameter direction, which defines the reaction-plane orientation in the transverse plane.
The decay angular distribution of the vector meson is therefore expected to exhibit a characteristic $\cos 2\Delta\phi$ modulation with respect to this plane, where $\Delta\phi$ denotes the azimuthal angle between a decay lepton and the reaction plane.
The strength of this modulation reflects the polarization--reaction-plane correlation, which is set by the spatial structure of the early electromagnetic field---a structure governed by the impact parameter and the nuclear size~\cite{Shao:2025oeb,Luo:2025ewj}.

Coherent photoproduction, initially observed in UPCs, has now been found to persist in peripheral and even semi-central hadronic heavy-ion collisions~\cite{ALICE:2015mzu,STAR:2019yox,LHCb:2021hoq,ALICE:2022zso}. In these collisions, the event plane reconstructed from the hadronic final state~\cite{Poskanzer:1998yz} provides an independent experimental estimate of the reaction plane, enabling a direct measurement of the vector-meson polarization relative to the collision geometry.
Among the vector mesons that could be considered for this measurement, the $J/\psi$ is particularly well suited.
Unlike the $\rho$, for which hadronic production remains dominant~\cite{Shen:2024eeb},
$J/\psi$ production at very low transverse momentum
($p_{\mathrm{T}} \lesssim 0.2~\mathrm{GeV}/c$) is dominated by photoproduction~\cite{STAR:2019yox,ALICE:2022zso,Klusek-Gawenda:2015hja,Zha:2018ytv,Zha:2017jch}.
This polarization--geometry correlation can provide an estimator of the reaction-plane direction in UPCs and low-multiplicity collisions, where conventional event-plane determination is unavailable or challenging.
Unlike conventional estimators based on final-state particle correlations, the polarization-based estimator is tied to the initial electromagnetic field and does not rely on final-state collective behavior.
Motivated by the opportunity to establish this polarization--geometry connection directly, we report in this Letter the first measurement of the decay angular distribution of photon-induced $J/\psi$ relative to the event plane in hadronic heavy-ion collisions.

This analysis uses data recorded in Ru+Ru and Zr+Zr collisions at a nucleon--nucleon center-of-mass energy of \(\sqrt{s_{\mathrm{NN}}} = 200~\mathrm{GeV}\) by the STAR experiment at RHIC in 2018~\cite{STAR:2002eio}. The two isobars have similar nuclear sizes, and their separately measured modulations agree within uncertainties; the data sets are therefore combined. The combined data set consists of approximately $3.5\times10^{9}$ minimum-bias events, selected by requiring coincident signals in the Vertex Position Detectors~\cite{Llope:2014vpd} covering the pseudorapidity range $4.24 < |\eta| < 5.1$. Collision centrality was determined from the charged-particle multiplicity measured with the Time Projection Chamber (TPC)~\cite{Anderson:2003ur} within $|\eta|<0.5$, using a Monte Carlo Glauber model~\cite{Miller:2007ri}. To avoid events affected by trigger inefficiency in the most peripheral region, the analysis is restricted to the 0--80\% centrality class, where 0\% denotes the most central collisions and larger percentiles correspond to more peripheral events.

$J/\psi$ candidates at midrapidity ($|y| < 0.8$) are reconstructed via the dielectron decay channel, $J/\psi \rightarrow e^{+}e^{-}$. Charged-particle tracks are measured by the TPC, with candidate tracks restricted to transverse momenta $p_{\mathrm{T}} > 0.5$ GeV/$c$ and pseudorapidity $|\eta| < 0.8$ to ensure uniform detector acceptance. Electron identification relies primarily on the ionization energy loss ($dE/dx$) in the TPC, expressed as the normalized standard deviation ($n\sigma_{e}$) from the theoretical Bichsel expectation~\cite{Bichsel:2006cs}. To effectively suppress hadronic contamination and ensure a high electron purity, tracks are required to be matched to at least one of the Time-of-Flight (TOF) and Barrel Electromagnetic Calorimeter (BEMC) detectors~\cite{Bonner:2003bv,STAR:2002ymp}, with TOF matching required for $p<1$ GeV/$c$, and to satisfy the corresponding electron-identification criterion based on the flight time or the energy-to-momentum ratio ($E/p$), respectively.

To quantify the polarization--geometry correlation of photoproduced $J/\psi$,
we measure the second-order azimuthal modulation of its decay electrons with respect to the
reconstructed event plane ($\Psi_{\mathrm{EP}}$). For the inclusive $J/\psi$ sample, the raw modulation is $A_2^{\rm raw}=\langle \cos 2(\phi_e-\Psi_{\mathrm{EP}})\rangle$, where $\phi_e$ is the electron azimuthal angle in the $J/\psi$ rest frame. Within each centrality interval, the modulation with respect to the reaction plane is obtained as $A_2^{\rm incl}=A_2^{\rm raw}/R_2$, where $R_2\equiv\mathrm{Res}\{\Psi_{\mathrm{EP}}\}$ is the ensemble-averaged second-order event-plane resolution.

The event plane $\Psi_{\mathrm{EP}}$ is reconstructed using the $p_{\mathrm{T}}$-weighted $Q$-vector method~\cite{Poskanzer:1998yz} with TPC tracks associated with the reconstructed primary vertex and satisfying $0.4 < p_{\mathrm{T}} < 3.0$ GeV/$c$. The second-order vector is defined as $\vec{Q}_2=\sum_i p_{\mathrm{T},i}(\cos 2\phi_i,\sin 2\phi_i)$, and the event-plane angle is $\Psi_{\mathrm{EP}}=\frac{1}{2}\operatorname{atan2}(Q_{2,y},Q_{2,x})$. To suppress self-correlations, electron tracks that can form a dielectron pair with an invariant mass in the range $2.2 < m_{e^{+}e^{-}} < 3.8$ GeV/$c^2$ are excluded from the event-plane reconstruction. The remaining TPC tracks are partitioned into east ($-1 < \eta < -0.05$) and west ($0.05 < \eta < 1$) subevents. Tracks near midrapidity ($|\eta|<0.05$), close to the TPC central membrane separating the two drift volumes, are excluded to avoid possible duplicate or split-track contributions. Standard recentering and shift corrections are applied in the event-plane angle determination to remove detector acceptance effects~\cite{Poskanzer:1998yz,Selyuzhenkov:2007zi}.
The residual Fourier components of the corrected event-plane distribution are below $10^{-3}$.
For the 30--80\% centrality class, the effective event-plane resolution is $R_2=0.46$.

The $J/\psi$ signal is obtained by subtracting the like-sign (LS, $e^{+}e^{+}$ and $e^{-}e^{-}$) combinatorial background from the unlike-sign (US, $e^{+}e^{-}$) invariant-mass distribution. Candidate pairs within the mass window $3.0 < m_{e^{+}e^{-}} < 3.2$ GeV/$c^2$ are selected for the modulation analysis. Each candidate is weighted by the inverse product of the single-electron efficiencies, $w = 1/(\epsilon_{e^{+}}\epsilon_{e^{-}})$. The TPC tracking efficiency is obtained from Monte Carlo simulations, while the TOF/BEMC matching and particle-identification efficiencies are evaluated with data-driven methods. The resulting $\Delta\phi = \phi_e - \Psi_{\mathrm{EP}}$ distribution is folded into the range $[0,\pi/2]$ and grouped into five intervals.
The raw modulation amplitude, $A_2^{\rm raw}$, is extracted from these intervals and corrected for the finite event-plane resolution to obtain the inclusive modulation before hadronic-background subtraction, $A_2^{\rm incl}$.

\begin{figure}[htb]
  \centering
  \begin{minipage}[b]{1.0\linewidth}
    \centering
    \includegraphics[width=\linewidth]{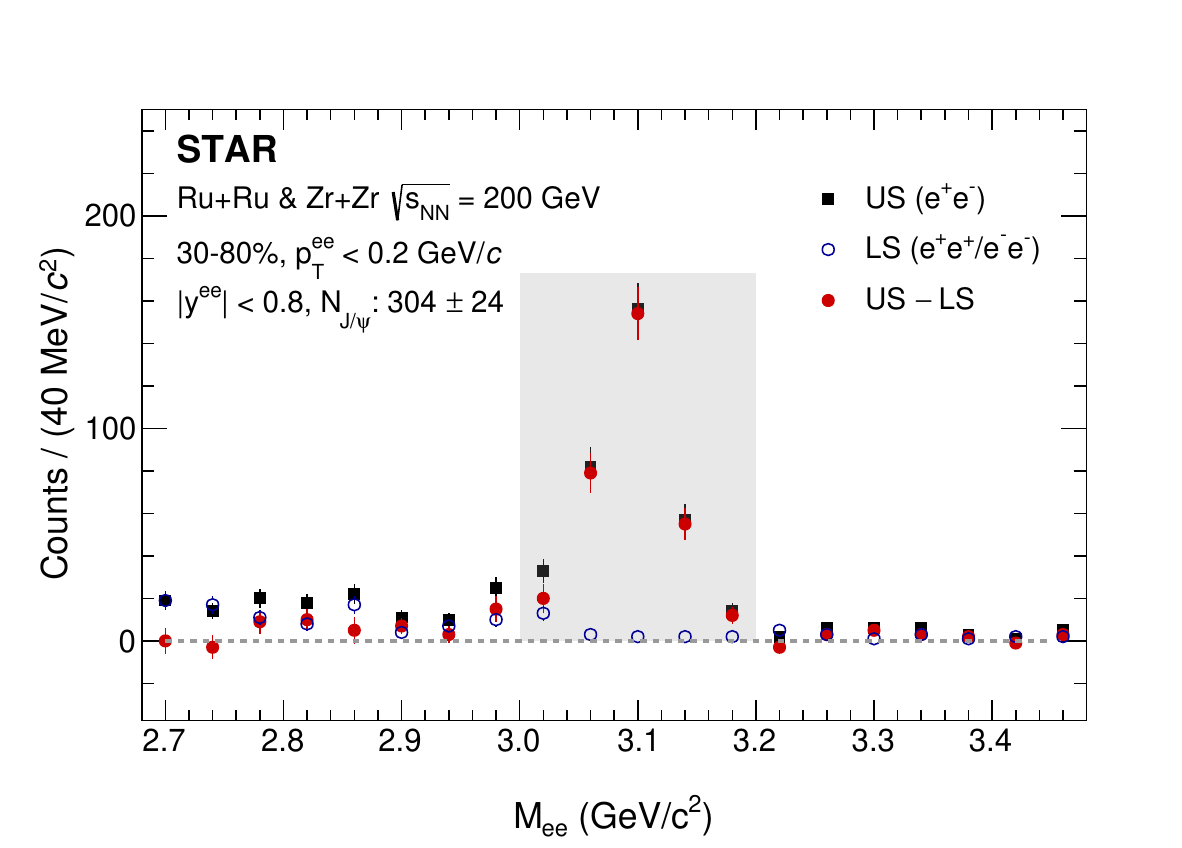}
  \end{minipage}
  \begin{minipage}[b]{1.0\linewidth}
    \centering
    \includegraphics[width=\linewidth]{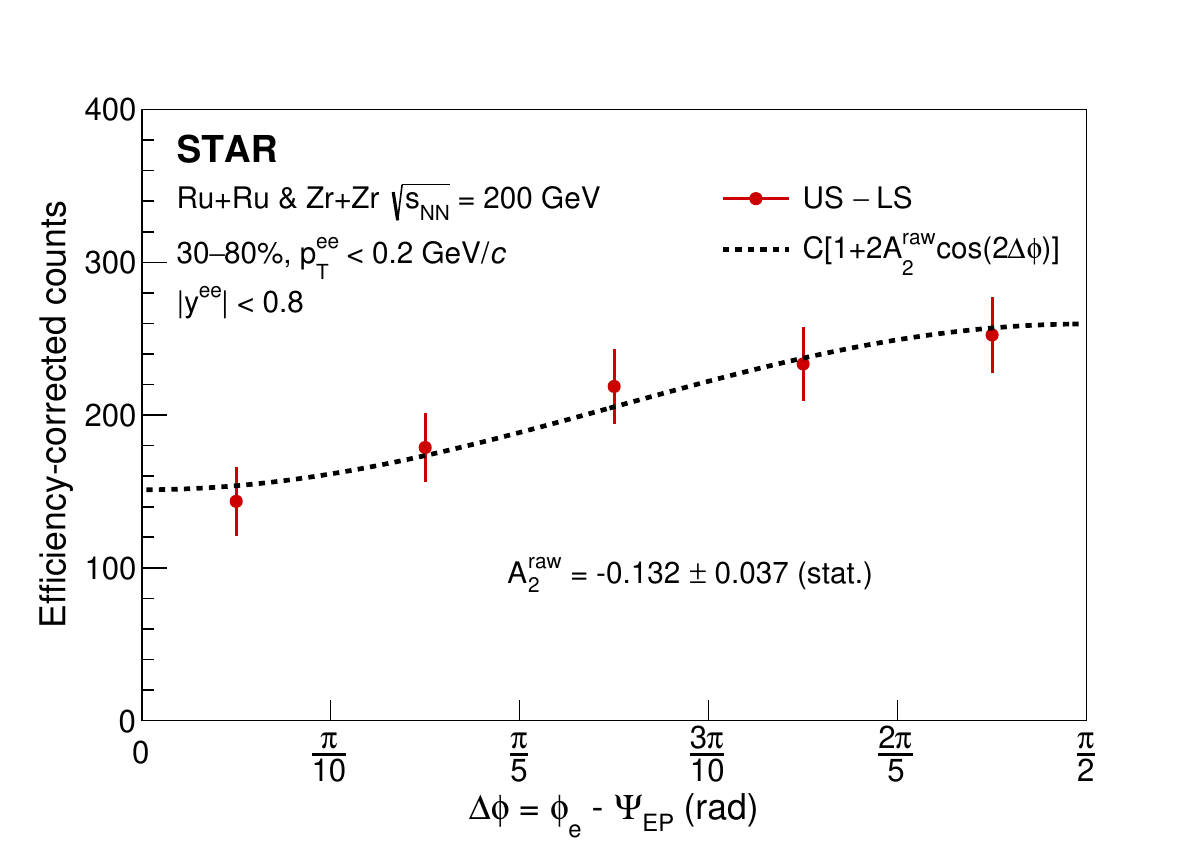}
  \end{minipage}
  \captionsetup{justification=raggedright,singlelinecheck=false,width=1\linewidth}
\caption{
Electron-pair invariant-mass distributions (upper) and the efficiency-corrected, like-sign-subtracted $\Delta\phi$ distribution (lower)
in Ru+Ru and Zr+Zr collisions at
$\sqrt{s_{\mathrm{NN}}}=200~\mathrm{GeV}$ for 30--80\% centrality,
with dielectron-pair rapidity $|y^{ee}|<0.8$ and $0<p_{\mathrm{T}}^{ee}<0.2~\mathrm{GeV}/c$.
In the upper panel, the shaded region marks the $J/\psi$ signal window.
In the lower panel, the dashed curve shows a fit with
$C[1+2A_2^{\mathrm{raw}}\cos(2\Delta\phi)]$.
}
  \label{fig:rawsignal}
\end{figure}

Figure~\ref{fig:rawsignal} shows the invariant-mass and efficiency-corrected $\Delta\phi$ distributions for
$e^+e^-$ pairs in the 30--80\% centrality class, within $|y^{ee}|<0.8$ and
$0<p_{\mathrm{T}}^{ee}<0.2~\mathrm{GeV}/c$.
In the upper panel, solid squares, open circles, and solid circles denote unlike-sign,
like-sign, and background-subtracted (US--LS) pairs, respectively.
A clear $J/\psi$ signal is observed, while residual contributions from QED processes
($\gamma\gamma\to e^+e^-$)~\cite{Zha:2020cst,Luo:2023syp} and the charm
continuum~\cite{STAR:2013pwb} are found to be negligible.
A fit to the lower-panel distribution with $C[1+2A_2^{\rm raw}\cos(2\Delta\phi)]$ gives
$A_2^{\rm raw}=-0.132\pm0.037$ (stat.). Possible non-flow correlations between the TPC subevents could increase the estimated event-plane resolution and thereby reduce the magnitude of the resolution-corrected $A_2$; such an effect cannot generate the observed raw modulation and would make the extracted signal more conservative. After the event-plane-resolution correction, the inclusive modulation is $A_2^{\rm incl}=-0.29\pm0.08$ (stat.) $\pm0.03$ (sys.).

Although the very-low-$p_{\mathrm{T}}^{ee}$ region ($p_{\mathrm{T}}^{ee}<0.2~\mathrm{GeV}/c$) is dominated by coherent photoproduction~\cite{STAR:2019yox}, the measured inclusive modulation, $A_2^{\rm incl}$, contains a residual hadronic contribution.
To estimate this contribution, the efficiency-corrected $J/\psi$ $p_{\mathrm{T}}^{ee}$ spectrum is constructed in each centrality interval.
The raw $J/\psi$ yields are extracted from invariant-mass fits with three components: a $J/\psi$ signal shape from Monte Carlo simulations~\cite{STAR:2019yox}, a mixed-event combinatorial background, and a linear residual background.
The extracted yields are then corrected for the total reconstruction efficiency, including TPC tracking, TOF/BEMC matching, and electron-identification efficiencies.
The hadronic yield fraction, $f_{\rm had}=Y_{\rm had}/Y_{\rm tot}$, is estimated by fitting the efficiency-corrected $J/\psi$ $p_{\mathrm{T}}^{ee}$ spectrum in the hadronic-dominated region $0.2 < p_{\mathrm{T}}^{ee} < 5.0~\mathrm{GeV}/c$ with a Tsallis function~\cite{Shao:2009mu,Tang:2008ud}, and extrapolating the fit to $p_{\mathrm{T}}^{ee}<0.2~\mathrm{GeV}/c$.
The photoproduction modulation is then evaluated as
\[
A_2^{\rm phot} = \frac{A_2^{\rm incl} - f_{\rm had} A_2^{\rm had}}{1 - f_{\rm had}} .
\]
Here, $A_2^{\rm had}$ denotes the modulation of the residual hadronic $J/\psi$ component and is taken from the measured modulation in $0.2 < p_{\mathrm{T}}^{ee} < 1.5~\mathrm{GeV}/c$. For the 30--80\% centrality class, the nominal value is $A_2^{\rm had}=-0.028\pm0.023$ (stat.) $\pm0.013$ (sys.).
For the 30--80\% centrality class, the extracted $f_{\rm had}$ values are $0.14 \pm 0.02$ (stat.) and $0.46 \pm 0.06$ (stat.) for $0 < p_{\mathrm{T}}^{ee} < 0.1~\mathrm{GeV}/c$ and $0.1 < p_{\mathrm{T}}^{ee} < 0.2~\mathrm{GeV}/c$, respectively, corresponding to an integrated fraction of $0.29 \pm 0.03$ (stat.) for $p_{\mathrm{T}}^{ee}<0.2~\mathrm{GeV}/c$.

\begin{figure*}[htb]
    \centering
     \begin{minipage}{\textwidth}
        \centering
        \includegraphics[width=\textwidth]{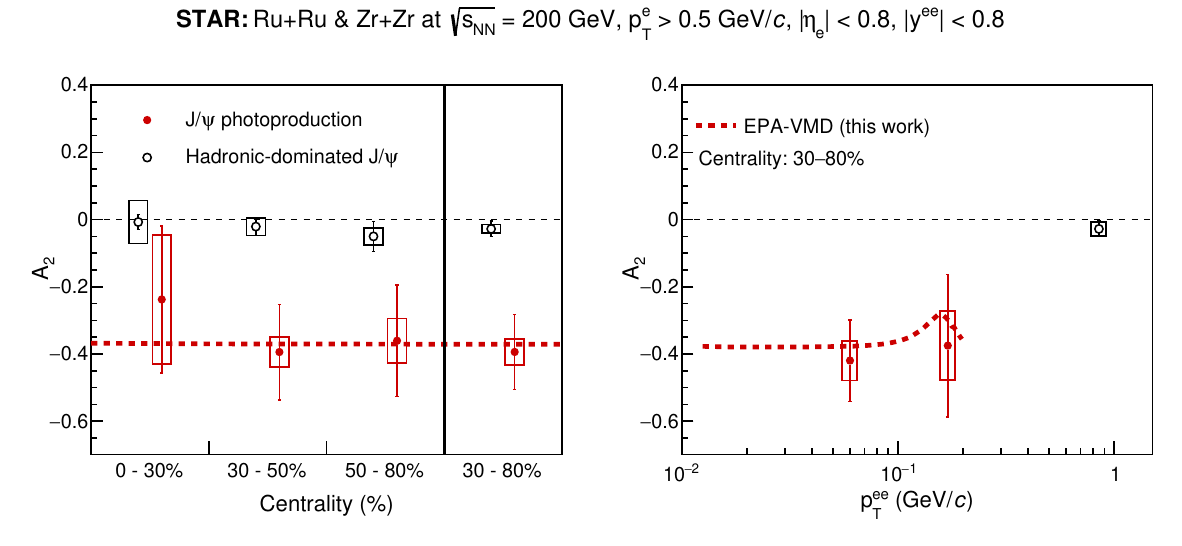}
    \end{minipage}
   \caption{
Second-order azimuthal modulation $A_2$ versus centrality (left)
and $p_{\mathrm{T}}^{ee}$ (right) in Ru+Ru and Zr+Zr collisions at
$\sqrt{s_{\mathrm{NN}}}=200$ GeV, measured within
$|\eta_e|<0.8$, $p_{\mathrm{T}}^e>0.5~\mathrm{GeV}/c$, and $|y^{ee}|<0.8$.
Solid and open circles correspond to the hadronic-background-subtracted photoproduction component for
$p_{\mathrm{T}}^{ee}<0.2~\mathrm{GeV}/c$ and the hadronic-dominated sample for
$0.2<p_{\mathrm{T}}^{ee}<1.5~\mathrm{GeV}/c$, respectively.
Vertical bars and boxes denote statistical and systematic uncertainties, respectively;
dashed lines show calculations based on the equivalent photon approximation and vector meson dominance (EPA--VMD).
}
    \label{fig:results}
\end{figure*}

Systematic uncertainties on the extracted photoproduction modulation, $A_2^{\rm phot}$, are evaluated in two main categories: signal extraction and hadronic background subtraction.
Uncertainties associated with signal extraction are estimated by varying the track selection criteria, electron identification requirements, and the invariant-mass window used for the modulation analysis.
Uncertainties arising from the hadronic background subtraction are evaluated by varying the invariant-mass fit ranges and background parameterizations, and by replacing the measured $A_2^{\rm had}$ with the assumption $A_2^{\rm had}=0$.
For the integrated 30--80\% result, tracking, electron identification, hadronic-background subtraction, and signal mass-window selection contribute 7\%, 7\%, 4\%, and 1\%, respectively.
Replacing the Tsallis fit with the empirical form $a/[1+b^2(p_{\mathrm{T}}^{ee})^2]^n$ used previously by STAR~\cite{STAR:2019yox} changes the low-$p_{\mathrm{T}}$ extrapolation negligibly.
Summing all independent contributions in quadrature, the total systematic uncertainty on $A_2^{\rm phot}$ for the integrated 30--80\% centrality class at $p_{\mathrm{T}}^{ee}<0.2$ GeV/$c$ is 0.04 (11\%).
For the differential measurements, the absolute systematic uncertainties range from 0.05 to 0.19 for the measured centrality classes, and from 0.06 to 0.10 for the measured $p_{\mathrm{T}}^{ee}$ intervals.

The final extracted second-order azimuthal modulations ($A_{2}^{\rm phot}$) for photoproduced
$J/\psi$ are summarized in Fig.~\ref{fig:results}.
The measurements are performed within $|\eta_{e}| < 0.8$, $p_{\mathrm{T}}^{e} > 0.5$ GeV/$c$,
and $|y^{ee}| < 0.8$ for both panels.
In both panels, solid circles represent the photoproduction component extracted
in the low-$p_{\mathrm{T}}^{ee}$ region after hadronic background subtraction, while open circles represent the hadronic-dominated sample at higher $p_{\mathrm{T}}^{ee}$.
These values are obtained by correcting the measured raw amplitudes for the finite
event-plane resolution and, for the solid circles, subsequently subtracting the hadronic
background contribution.

The left panel displays the centrality dependence of $A_{2}$ in Ru+Ru and Zr+Zr
collisions at $\sqrt{s_{\mathrm{NN}}}=200$ GeV.
The central values of $A_2^{\rm phot}$ are negative across the measured centrality
intervals in the photoproduction-dominated region ($p_{\mathrm{T}}^{ee} < 0.2$
GeV/$c$), although the most central interval has a larger uncertainty.
For the combined 30--80\% centrality class, the photoproduction component is
determined to be
$A_{2}^{\rm phot} = -0.39 \pm 0.11 \text{(stat.)} \pm 0.04 \text{(sys.)}$.
The negative sign reflects the $J/\psi\to e^+e^-$ decay structure, in which
spin-$1/2$ leptons are preferentially emitted perpendicular to the $J/\psi$
polarization and hence to the event-plane direction.

The right panel of Fig.~\ref{fig:results} shows the $p_{\mathrm{T}}^{ee}$ dependence
for the 30--80\% centrality class.
The combined 30--80\% result for $p_{\mathrm{T}}^{ee} < 0.2$ GeV/$c$
is significantly below zero, providing evidence
for linear polarization of coherently photoproduced $J/\psi$.
Any potential $p_{\mathrm{T}}^{ee}$ dependence arising from the interplay between
the photon's linear polarization and the nuclear diffractive structure~\cite{Brandenburg:2022jgr,STAR:2022wfe}
remains unresolved within the present uncertainties.
The dominance of the coherent process in this very-low-$p_{\mathrm{T}}^{ee}$ regime
ensures that the modulation is not significantly diluted by incoherent scattering,
which is sensitive to nucleon-level fluctuations and nuclear breakup.
Instead, the coherent interaction intrinsically probes the bulk nuclear geometry,
preserving a robust alignment between the $J/\psi$ spin and the reaction-plane
geometry. This provides a clean signature of the polarization--geometry correlation carried by the initial electromagnetic fields.

The data are compared with EPA--VMD calculations following Ref.~\cite{Wu:2022exl}.
The model provides a good description of the measured $A_2^{\rm phot}$ across both centrality and $p_{\mathrm{T}}$ intervals, supporting the interpretation that the observed negative modulation originates from the decay distribution of a linearly polarized $J/\psi$.
The measured $A_2^{\rm phot}$ exhibits no significant centrality dependence, a behavior that is also consistent with the model baseline across the presented centrality intervals.
In the Good--Walker picture, coherent production probes the configuration-averaged nuclear geometry, while the photon-polarization axis is tied to the impact-parameter direction by the initial electromagnetic field~\cite{Good:1960ba,Lappi:2010dd,Wu:2022exl}.
The resulting polarization-based observable is therefore intrinsically insensitive to final-state non-flow correlations.

In summary, we have reported the first event-plane-resolved measurement of the decay angular distribution of photon-induced $J/\psi$ in hadronic Ru+Ru and Zr+Zr collisions at $\sqrt{s_{\mathrm{NN}}}=200$ GeV. For $p_{\mathrm{T}}^{ee}<0.2~\mathrm{GeV}/c$ in the 30--80\% centrality class, the extracted photoproduction modulation is $A_2^{\rm phot}=-0.39\pm0.11$ (stat.) $\pm0.04$ (sys.), providing evidence for a negative second-order modulation consistent with linearly polarized photoproduction. This observation demonstrates that the produced $J/\psi$ retains a strong correlation with the reaction-plane direction and establishes polarized photoproduction as a direct probe of the transverse orientation of the initial collision geometry. With larger data samples, this polarization-based reference could enable geometry-sensitive measurements in small or low-multiplicity systems.

\bigskip

We thank the RHIC Operations Group and SCDF at BNL, the NERSC Center at LBNL, and the Open Science Grid consortium for providing resources and support.  This work was supported in part by the Office of Nuclear Physics within the U.S. DOE Office of Science, the U.S. National Science Foundation, National Natural Science Foundation of China, Chinese Academy of Science, the Ministry of Science and Technology of China and the Chinese Ministry of Education, NSTC Taipei, the National Research Foundation of Korea, Czech Science Foundation and Ministry of Education, Youth and Sports of the Czech Republic, Hungarian National Research, Development and Innovation Office, New National Excellency Programme of the Hungarian Ministry of Human Capacities, Department of Atomic Energy and Department of Science and Technology of the Government of India, the National Science Centre and WUT ID-UB of Poland, German Bundesministerium f\"ur Bildung, Wissenschaft, Forschung and Technologie (BMBF), Helmholtz Association, Ministry of Education, Culture, Sports, Science, and Technology (MEXT), Japan Society for the Promotion of Science (JSPS), and Agencia Nacional de Investigacion y Desarrollo de Chile (ANID), Chile.

\bibliography{references}

@article{Shao:2009mu,
    author = "Shao, Ming and Yi, Li and Tang, Zebo and Chen, Hongfang and Li, Cheng and Xu, Zhangbu",
    title = "{Examine the species and beam-energy dependence of particle spectra using Tsallis Statistics}",
    eprint = "0912.0993",
    archivePrefix = "arXiv",
    primaryClass = "nucl-ex",
    doi = "10.1088/0954-3899/37/8/085104",
    journal = "J. Phys. G",
    volume = "37",
    pages = "085104",
    year = "2010"
}

@article{Bichsel:2006cs,
    author = "Bichsel, H.",
    title = "{A method to improve tracking and particle identification in TPCs and silicon detectors}",
    doi = "10.1016/j.nima.2006.03.009",
    journal = "Nucl. Instrum. Meth. A",
    volume = "562",
    pages = "154--197",
    year = "2006"
}

@article{Wu:2025dxg,
    author = "Wu, Xin and Li, Xin-Bai and Tang, Ze-Bo and Wang, Kai-Yang and Zha, Wang-Mei",
    title = "{Impact parameter manipulation in exclusive photoproduction in Electron-Ion Collisions}",
    doi = "10.1007/s41365-025-01704-5",
    journal = "Nucl. Sci. Tech.",
    volume = "36",
    number = "9",
    pages = "157",
    year = "2025"
}

@article{Huang:2015oca,
    author = "Huang, Xu-Guang",
    title = "{Electromagnetic fields and anomalous transports in heavy-ion collisions --- A pedagogical review}",
    eprint = "1509.04073",
    archivePrefix = "arXiv",
    primaryClass = "nucl-th",
    doi = "10.1088/0034-4885/79/7/076302",
    journal = "Rept. Prog. Phys.",
    volume = "79",
    number = "7",
    pages = "076302",
    year = "2016"
}

@article{Shen:2025unr,
    author = "Shen, Diyu and Chen, Jinhui and Huang, Xu-Guang and Ma, Yu-Gang and Tang, Aihong and Wang, Gang",
    title = "{A Review of Intense Electromagnetic Fields in Heavy-Ion Collisions: Theoretical Predictions and Experimental Results}",
    eprint = "2512.00739",
    archivePrefix = "arXiv",
    primaryClass = "nucl-ex",
    doi = "10.34133/research.0726",
    journal = "Research",
    volume = "8",
    pages = "0726",
    year = "2025"
}

@article{Shao:2025oeb,
    author = "Shao, Ding Yu and Yu, Han-Qing and Zhang, Cheng and Zhou, Jian",
    title = "{Geometry-induced azimuthal anisotropy in coherent J/{\ensuremath{\psi}} photoproduction}",
    eprint = "2511.17670",
    archivePrefix = "arXiv",
    primaryClass = "hep-ph",
    doi = "10.1103/fhks-8s8z",
    journal = "Phys. Rev. D",
    volume = "113",
    number = "9",
    pages = "094022",
    year = "2026"
}

@article{Li:2019sin,
    author = "Li, Cong and Zhou, Jian and Zhou, Ya-Jin",
    title = "{Impact parameter dependence of the azimuthal asymmetry in lepton pair production in heavy ion collisions}",
    eprint = "1911.00237",
    archivePrefix = "arXiv",
    primaryClass = "hep-ph",
    doi = "10.1103/PhysRevD.101.034015",
    journal = "Phys. Rev. D",
    volume = "101",
    number = "3",
    pages = "034015",
    year = "2020"
}

@article{Xiao:2020ddm,
    author = "Xiao, Bo-Wen and Yuan, Feng and Zhou, Jian",
    title = "{Momentum Anisotropy of Leptons from Two Photon Processes in Heavy Ion Collisions}",
    eprint = "2003.06352",
    archivePrefix = "arXiv",
    primaryClass = "hep-ph",
    doi = "10.1103/PhysRevLett.125.232301",
    journal = "Phys. Rev. Lett.",
    volume = "125",
    number = "23",
    pages = "232301",
    year = "2020"
}

@article{Good:1960ba,
    author = "Good, M. L. and Walker, W. D.",
    title = "{Diffraction dissociation of beam particles}",
    doi = "10.1103/PhysRev.120.1857",
    journal = "Phys. Rev.",
    volume = "120",
    pages = "1857--1860",
    year = "1960"
}

@article{Zha:2018tlq,
    author = "Zha, Wangmei and Brandenburg, James Daniel and Tang, Zebo and Xu, Zhangbu",
    title = "{Initial transverse-momentum broadening of Breit-Wheeler process in relativistic heavy-ion collisions}",
    eprint = "1812.02820",
    archivePrefix = "arXiv",
    primaryClass = "nucl-th",
    doi = "10.1016/j.physletb.2019.135089",
    journal = "Phys. Lett. B",
    volume = "800",
    pages = "135089",
    year = "2020"
}

@article{Li:2023yjt,
    author = "Li, Xinbai. and Luo, Jiaxuan. and Tang, Zebo. and Wu, Xin. and Zha, Wangmei.",
    title = "{Exploring the higher-order QED effects on the differential distributions of vacuum pair production in relativistic heavy-ion collisions}",
    eprint = "2307.01549",
    archivePrefix = "arXiv",
    primaryClass = "hep-ph",
    doi = "10.1016/j.physletb.2023.138314",
    journal = "Phys. Lett. B",
    volume = "847",
    pages = "138314",
    year = "2023"
}

@article{Klusek-Gawenda:2016nuo,
    author = {K{\l}usek-Gawenda, Mariola and Sch{\"a}fer, Wolfgang and Szczurek, Antoni},
    title = "{Two-gluon exchange contribution to elastic $\gamma \gamma \to \gamma \gamma$ scattering and production of two-photons in ultraperipheral ultrarelativistic heavy ion and proton-proton collisions}",
    eprint = "1606.01058",
    archivePrefix = "arXiv",
    primaryClass = "hep-ph",
    doi = "10.1016/j.physletb.2016.08.059",
    journal = "Phys. Lett. B",
    volume = "761",
    pages = "399--407",
    year = "2016"
}

@article{Baltz:2007gs,
    author = "Baltz, A. J.",
    title = "{Evidence for higher order QED in e+ e- pair production at RHIC}",
    eprint = "0710.4944",
    archivePrefix = "arXiv",
    primaryClass = "nucl-th",
    doi = "10.1103/PhysRevLett.100.062302",
    journal = "Phys. Rev. Lett.",
    volume = "100",
    pages = "062302",
    year = "2008"
}

@article{Tang:2008ud,
    author = "Tang, Zebo and Xu, Yichun and Ruan, Lijuan and van Buren, Gene and Wang, Fuqiang and Xu, Zhangbu",
    title = "{Spectra and radial flow at RHIC with Tsallis statistics in a Blast-Wave description}",
    eprint = "0812.1609",
    archivePrefix = "arXiv",
    primaryClass = "nucl-ex",
    reportNumber = "BNL-KB-02-02",
    doi = "10.1103/PhysRevC.79.051901",
    journal = "Phys. Rev. C",
    volume = "79",
    pages = "051901",
    year = "2009"
}

@article{STAR:2002eio,
    author = "Ackermann, K. H. and others",
    collaboration = "STAR",
    title = "{STAR detector overview}",
    doi = "10.1016/S0168-9002(02)01960-5",
    journal = "Nucl. Instrum. Meth. A",
    volume = "499",
    pages = "624--632",
    year = "2003"
}

@article{Luo:2025ewj,
    author = "Luo, Jiaxuan and Li, Xinbai and Tang, Zebo and Wu, Xin and Yang, Shuai and Zha, Wangmei and Zhang, Zhan",
    title = "{Probing the Collision Geometry via Two-Photon Processes in Heavy-Ion Collisions}",
    eprint = "2505.05133",
    archivePrefix = "arXiv",
    primaryClass = "hep-ph",
    doi = "10.1007/s41365-026-01934-1",
    journal = "Nucl. Sci. Tech.",
    volume = "37",
    number = "6",
    pages = "102",
    year = "2026"
}

@article{Anderson:2003ur,
    author = "Anderson, M. and others",
    title = "{The Star time projection chamber: A Unique tool for studying high multiplicity events at RHIC}",
    eprint = "nucl-ex/0301015",
    archivePrefix = "arXiv",
    doi = "10.1016/S0168-9002(02)01964-2",
    journal = "Nucl. Instrum. Meth. A",
    volume = "499",
    pages = "659--678",
    year = "2003"
}

@article{Bonner:2003bv,
    author = "Bonner, B. and Chen, H. and Eppley, G. and Geurts, F. and Lamas Valverde, J. and Li, C. and Llope, W. J. and Nussbaum, T. and Platner, E. and Roberts, J.",
    editor = "Fonte, P. and Fraga, M. and Ratti, S. P. and Santonico, R.",
    title = "{A single Time-of-Flight tray based on multigap resistive plate chambers for the STAR experiment at RHIC}",
    doi = "10.1016/S0168-9002(03)01347-0",
    journal = "Nucl. Instrum. Meth. A",
    volume = "508",
    pages = "181--184",
    year = "2003"
}

@article{STAR:2002ymp,
    author = "Beddo, M. and others",
    collaboration = "STAR",
    title = "{The STAR barrel electromagnetic calorimeter}",
    doi = "10.1016/S0168-9002(02)01970-8",
    journal = "Nucl. Instrum. Meth. A",
    volume = "499",
    pages = "725--739",
    year = "2003"
}

@article{Miller:2007ri,
    author = "Miller, Michael L. and Reygers, Klaus and Sanders, Stephen J. and Steinberg, Peter",
    title = "{Glauber modeling in high energy nuclear collisions}",
    eprint = "nucl-ex/0701025",
    archivePrefix = "arXiv",
    doi = "10.1146/annurev.nucl.57.090506.123020",
    journal = "Ann. Rev. Nucl. Part. Sci.",
    volume = "57",
    pages = "205--243",
    year = "2007"
}

@article{Braun-Munzinger:2007edi,
    author = "Braun-Munzinger, Peter and Stachel, Johanna",
    title = "{The quest for the quark-gluon plasma}",
    doi = "10.1038/nature06080",
    journal = "Nature",
    volume = "448",
    pages = "302--309",
    year = "2007"
}

@article{Skokov:2009qp,
    author = "Skokov, V. and Illarionov, A. Yu. and Toneev, V.",
    title = "{Estimate of the magnetic field strength in heavy-ion collisions}",
    eprint = "0907.1396",
    archivePrefix = "arXiv",
    primaryClass = "nucl-th",
    doi = "10.1142/S0217751X09047570",
    journal = "Int. J. Mod. Phys. A",
    volume = "24",
    pages = "5925--5932",
    year = "2009"
}

@article{Li:2019yzy,
    author = "Li, Cong and Zhou, Jian and Zhou, Ya-Jin",
    title = "{Probing the linear polarization of photons in ultraperipheral heavy ion collisions}",
    eprint = "1903.10084",
    archivePrefix = "arXiv",
    primaryClass = "hep-ph",
    doi = "10.1016/j.physletb.2019.07.005",
    journal = "Phys. Lett. B",
    volume = "795",
    pages = "576--580",
    year = "2019"
}

@article{ALICE:2022zso,
    author = "Acharya, Shreyasi and others",
    collaboration = "ALICE",
    title = "{Photoproduction of low-$p_{\rm T}$ J/$\psi$ from peripheral to central Pb$-$Pb collisions at 5.02 TeV}",
    eprint = "2204.10684",
    archivePrefix = "arXiv",
    primaryClass = "nucl-ex",
    reportNumber = "CERN-EP-2022-071",
    doi = "10.1016/j.physletb.2022.137467",
    journal = "Phys. Lett. B",
    volume = "846",
    pages = "137467",
    year = "2023"
}

@article{Hattori:2016emy,
    author = "Hattori, Koichi and Huang, Xu-Guang",
    title = "{Novel quantum phenomena induced by strong magnetic fields in heavy-ion collisions}",
    eprint = "1609.00747",
    archivePrefix = "arXiv",
    primaryClass = "nucl-th",
    reportNumber = "RBRC-1202",
    doi = "10.1007/s41365-016-0178-3",
    journal = "Nucl. Sci. Tech.",
    volume = "28",
    number = "2",
    pages = "26",
    year = "2017"
}

@article{Deng:2012pc,
    author = "Deng, Wei-Tian and Huang, Xu-Guang",
    title = "{Event-by-event generation of electromagnetic fields in heavy-ion collisions}",
    eprint = "1201.5108",
    archivePrefix = "arXiv",
    primaryClass = "nucl-th",
    doi = "10.1103/PhysRevC.85.044907",
    journal = "Phys. Rev. C",
    volume = "85",
    pages = "044907",
    year = "2012"
}

@article{Bertulani:2005ru,
    author = "Bertulani, Carlos A. and Klein, Spencer R. and Nystrand, Joakim",
    title = "{Physics of ultra-peripheral nuclear collisions}",
    eprint = "nucl-ex/0502005",
    archivePrefix = "arXiv",
    doi = "10.1146/annurev.nucl.55.090704.151526",
    journal = "Ann. Rev. Nucl. Part. Sci.",
    volume = "55",
    pages = "271--310",
    year = "2005"
}

@article{Baur:2001jj,
    author = "Baur, Gerhard and Hencken, Kai and Trautmann, Dirk and Sadovsky, Serguei and Kharlov, Yuri",
    title = "{Coherent gamma gamma and gamma-A interactions in very peripheral collisions at relativistic ion colliders}",
    eprint = "hep-ph/0112211",
    archivePrefix = "arXiv",
    doi = "10.1016/S0370-1573(01)00101-6",
    journal = "Phys. Rept.",
    volume = "364",
    pages = "359--450",
    year = "2002"
}

@article{Klein:2020fmr,
    author = "Klein, Spencer and Steinberg, Peter",
    title = "{Photonuclear and Two-photon Interactions at High-Energy Nuclear Colliders}",
    eprint = "2005.01872",
    archivePrefix = "arXiv",
    primaryClass = "nucl-ex",
    doi = "10.1146/annurev-nucl-030320-033923",
    journal = "Ann. Rev. Nucl. Part. Sci.",
    volume = "70",
    pages = "323--354",
    year = "2020"
}

@article{Klein:1999qj,
    author = "Klein, Spencer and Nystrand, Joakim",
    title = "{Exclusive vector meson production in relativistic heavy ion collisions}",
    eprint = "hep-ph/9902259",
    archivePrefix = "arXiv",
    reportNumber = "LBNL-42768, LBL-42768",
    doi = "10.1103/PhysRevC.60.014903",
    journal = "Phys. Rev. C",
    volume = "60",
    pages = "014903",
    year = "1999"
}

@article{Toll:2012mb,
    author = "Toll, Tobias and Ullrich, Thomas",
    title = "{Exclusive diffractive processes in electron-ion collisions}",
    eprint = "1211.3048",
    archivePrefix = "arXiv",
    primaryClass = "hep-ph",
    doi = "10.1103/PhysRevC.87.024913",
    journal = "Phys. Rev. C",
    volume = "87",
    number = "2",
    pages = "024913",
    year = "2013"
}

@article{Guzey:2013xba,
    author = "Guzey, V. and Kryshen, E. and Strikman, M. and Zhalov, M.",
    title = "{Evidence for nuclear gluon shadowing from the ALICE measurements of PbPb ultraperipheral exclusive $J/{\psi}$ production}",
    eprint = "1305.1724",
    archivePrefix = "arXiv",
    primaryClass = "hep-ph",
    doi = "10.1016/j.physletb.2013.08.043",
    journal = "Phys. Lett. B",
    volume = "726",
    pages = "290--295",
    year = "2013"
}

@article{Wu:2022exl,
    author = "Wu, Xin and Li, Xinbai and Tang, Zebo and Wang, Pengfei and Zha, Wangmei",
    title = "{Reaction plane alignment with linearly polarized photon in heavy-ion collisions}",
    eprint = "2302.10458",
    archivePrefix = "arXiv",
    primaryClass = "hep-ph",
    doi = "10.1103/PhysRevResearch.4.L042048",
    journal = "Phys. Rev. Res.",
    volume = "4",
    number = "4",
    pages = "L042048",
    year = "2022"
}

@article{Xing:2020hwh,
    author = "Xing, Hongxi and Zhang, Cheng and Zhou, Jian and Zhou, Ya-Jin",
    title = "{The cos 2$\phi$ azimuthal asymmetry in $\rho^{0}$ meson production in ultraperipheral heavy ion collisions}",
    eprint = "2006.06206",
    archivePrefix = "arXiv",
    primaryClass = "hep-ph",
    doi = "10.1007/JHEP10(2020)064",
    journal = "JHEP",
    volume = "10",
    pages = "064",
    year = "2020"
}

@article{Brandenburg:2022jgr,
    author = "Brandenburg, James Daniel and Xu, Zhangbu and Zha, Wangmei and Zhang, Cheng and Zhou, Jian and Zhou, Yajin",
    title = "{Exploring gluon tomography with polarization dependent diffractive J/{\ensuremath{\psi}} production}",
    eprint = "2207.02478",
    archivePrefix = "arXiv",
    primaryClass = "hep-ph",
    doi = "10.1103/PhysRevD.106.074008",
    journal = "Phys. Rev. D",
    volume = "106",
    number = "7",
    pages = "074008",
    year = "2022"
}

@article{Klein:1999gv,
    author = "Klein, Spencer R. and Nystrand, Joakim",
    title = "{Interference in exclusive vector meson production in heavy ion collisions}",
    eprint = "hep-ph/9909237",
    archivePrefix = "arXiv",
    reportNumber = "LBNL-43907, LBL-43907, LBNL-PUB-43907, LBNL-PUB-43907",
    doi = "10.1103/PhysRevLett.84.2330",
    journal = "Phys. Rev. Lett.",
    volume = "84",
    pages = "2330--2333",
    year = "2000"
}

@article{STAR:2022wfe,
    author = "Abdallah, Mohamed and others",
    collaboration = "STAR",
    title = "{Tomography of ultrarelativistic nuclei with polarized photon-gluon collisions}",
    eprint = "2204.01625",
    archivePrefix = "arXiv",
    primaryClass = "nucl-ex",
    doi = "10.1126/sciadv.abq3903",
    journal = "Sci. Adv.",
    volume = "9",
    number = "1",
    pages = "eabq3903",
    year = "2023"
}

@article{STAR:2025wpi,
    author = "Aboona, B. E. and others",
    collaboration = "STAR",
    title = "{Evidence of Spin-Interference Effects in Exclusive $J/\psi\to e^+e^-$ Photoproduction in Ultraperipheral Heavy-Ion Collisions}",
    eprint = "2512.02865",
    archivePrefix = "arXiv",
    primaryClass = "nucl-ex",
    doi = "10.1103/tcdb-ldh8",
    journal = "Phys. Rev. Lett.",
    volume = "136",
    number = "24",
    pages = "242302",
    year = "2026"
}

@article{ALICE:2015mzu,
    author = "Adam, Jaroslav and others",
    collaboration = "ALICE",
    title = "{Measurement of an excess in the yield of $J/\psi$ at very low $p_{\rm T}$ in Pb-Pb collisions at $\sqrt{s_{\rm NN}}$ = 2.76 TeV}",
    eprint = "1509.08802",
    archivePrefix = "arXiv",
    primaryClass = "nucl-ex",
    reportNumber = "CERN-PH-EP-2015-268",
    doi = "10.1103/PhysRevLett.116.222301",
    journal = "Phys. Rev. Lett.",
    volume = "116",
    number = "22",
    pages = "222301",
    year = "2016"
}

@article{ALICE:2024aza,
    author = "Acharya, Shreyasi and others",
    collaboration = "ALICE",
    title = "{Measurement of the impact-parameter dependent azimuthal anisotropy in coherent $\rho^0$ photoproduction in Pb$-$Pb collisions at $\sqrt{s_{\rm NN}}$ = 5.02 TeV}",
    eprint = "2405.14525",
    archivePrefix = "arXiv",
    primaryClass = "nucl-ex",
    doi = "10.1016/j.physletb.2024.139017",
    journal = "Phys. Lett. B",
    volume = "858",
    pages = "139017",
    year = "2024"
}

@article{STAR:2019wlg,
    author = "Adam, Jaroslav and others",
    collaboration = "STAR",
    title = "{Measurement of $e^+e^-$ Momentum and Angular Distributions from Linearly Polarized Photon Collisions}",
    eprint = "1910.12400",
    archivePrefix = "arXiv",
    primaryClass = "nucl-ex",
    doi = "10.1103/PhysRevLett.127.052302",
    journal = "Phys. Rev. Lett.",
    volume = "127",
    number = "5",
    pages = "052302",
    year = "2021"
}

@article{Zha:2017jch,
    author = "Zha, W. and Klein, S. R. and Ma, R. and Ruan, L. and Todoroki, T. and Tang, Z. and Xu, Z. and Yang, C. and Yang, Q. and Yang, S.",
    title = "{Coherent J/$\psi$ photoproduction in hadronic heavy-ion collisions}",
    eprint = "1705.01460",
    archivePrefix = "arXiv",
    primaryClass = "nucl-th",
    doi = "10.1103/PhysRevC.97.044910",
    journal = "Phys. Rev. C",
    volume = "97",
    number = "4",
    pages = "044910",
    year = "2018"
}

@article{STAR:2013pwb,
    author = "Adamczyk, L. and others",
    collaboration = "STAR",
    title = "{Dielectron Mass Spectra from Au+Au Collisions at $\sqrt{s_{\rm NN}}$ = 200 GeV}",
    eprint = "1312.7397",
    archivePrefix = "arXiv",
    primaryClass = "hep-ex",
    doi = "10.1103/PhysRevLett.113.022301",
    journal = "Phys. Rev. Lett.",
    volume = "113",
    number = "2",
    pages = "022301",
    year = "2014",
    note = "[Addendum: Phys.Rev.Lett. 113, 049903 (2014)]"
}

@article{Luo:2023syp,
    author = "Luo, Jiaxuan. and Li, Xinbai. and Tang, Zebo. and Wu, Xin. and Zha, Wangmei.",
    title = "{Effect of initial nuclear deformation on dielectron photoproduction in hadronic heavy-ion collisions}",
    eprint = "2308.03070",
    archivePrefix = "arXiv",
    primaryClass = "hep-ph",
    doi = "10.1103/PhysRevC.108.054906",
    journal = "Phys. Rev. C",
    volume = "108",
    number = "5",
    pages = "054906",
    year = "2023"
}

@article{Klusek-Gawenda:2015hja,
    author = "K{\l}usek-Gawenda, Mariola and Szczurek, Antoni",
    title = "{Photoproduction of $J/\psi$ mesons in peripheral and semicentral heavy ion collisions}",
    eprint = "1509.03173",
    archivePrefix = "arXiv",
    primaryClass = "nucl-th",
    doi = "10.1103/PhysRevC.93.044912",
    journal = "Phys. Rev. C",
    volume = "93",
    number = "4",
    pages = "044912",
    year = "2016"
}

@article{STAR:2019yox,
    author = "Adam, J. and others",
    collaboration = "STAR",
    title = "{Observation of excess J/$\psi$ yield at very low transverse momenta in Au+Au collisions at $\sqrt{s_{\rm{NN}}} =$ 200 GeV and U+U collisions at $\sqrt{s_{\rm{NN}}} =$ 193 GeV}",
    eprint = "1904.11658",
    archivePrefix = "arXiv",
    primaryClass = "hep-ex",
    doi = "10.1103/PhysRevLett.123.132302",
    journal = "Phys. Rev. Lett.",
    volume = "123",
    number = "13",
    pages = "132302",
    year = "2019"
}

@article{Zha:2020cst,
    author = "Zha, Wangmei and Brandenburg, James Daniel and Ruan, Lijuan and Tang, Zebo and Xu, Zhangbu",
    title = "{Exploring the double-slit interference with linearly polarized photons}",
    eprint = "2006.12099",
    archivePrefix = "arXiv",
    primaryClass = "hep-ph",
    doi = "10.1103/PhysRevD.103.033007",
    journal = "Phys. Rev. D",
    volume = "103",
    number = "3",
    pages = "033007",
    year = "2021"
}

@article{Zha:2018ytv,
    author = "Zha, W. and Ruan, L. and Tang, Z. and Xu, Z. and Yang, S.",
    title = "{Coherent photo-produced J$/\psi$ and dielectron yields in isobaric collisions}",
    eprint = "1810.02064",
    archivePrefix = "arXiv",
    primaryClass = "hep-ph",
    doi = "10.1016/j.physletb.2018.12.041",
    journal = "Phys. Lett. B",
    volume = "789",
    pages = "238--242",
    year = "2019"
}

@article{Shen:2024eeb,
    author = "Shen, Kaifeng and Wu, Xin and Tang, Zebo and Zha, Wangmei",
    title = "{Exploring the photoproduction of $\rho $ and $\phi $ in hadronic heavy-ion collisions}",
    eprint = "2401.07480",
    archivePrefix = "arXiv",
    primaryClass = "hep-ph",
    doi = "10.1140/epjc/s10052-024-13503-0",
    journal = "Eur. Phys. J. C",
    volume = "84",
    number = "11",
    pages = "1167",
    year = "2024"
}

@article{Poskanzer:1998yz,
    author = "Poskanzer, Arthur M. and Voloshin, S. A.",
    title = "{Methods for analyzing anisotropic flow in relativistic nuclear collisions}",
    eprint = "nucl-ex/9805001",
    archivePrefix = "arXiv",
    doi = "10.1103/PhysRevC.58.1671",
    journal = "Phys. Rev. C",
    volume = "58",
    pages = "1671--1678",
    year = "1998"
}

@article{Brandenburg:2021lnj,
    author = "Brandenburg, James Daniel and Zha, Wangmei and Xu, Zhangbu",
    title = "{Mapping the electromagnetic fields of heavy-ion collisions with the Breit-Wheeler process}",
    eprint = "2103.16623",
    archivePrefix = "arXiv",
    primaryClass = "hep-ph",
    doi = "10.1140/epja/s10050-021-00595-5",
    journal = "Eur. Phys. J. A",
    volume = "57",
    number = "10",
    pages = "299",
    year = "2021"
}

@article{Lappi:2010dd,
    author = "Lappi, T. and Mantysaari, H.",
    title = "{Incoherent diffractive J/Psi-production in high energy nuclear DIS}",
    eprint = "1011.1988",
    archivePrefix = "arXiv",
    primaryClass = "hep-ph",
    reportNumber = "INT-PUB-10-061",
    doi = "10.1103/PhysRevC.83.065202",
    journal = "Phys. Rev. C",
    volume = "83",
    pages = "065202",
    year = "2011"
}

@article{Adeluyi:2012ph,
    author = "Adeluyi, Adeola and Bertulani, C. A.",
    title = "{Constraining Gluon Shadowing Using Photoproduction in Ultraperipheral pA and AA Collisions}",
    eprint = "1201.0146",
    archivePrefix = "arXiv",
    primaryClass = "nucl-th",
    doi = "10.1103/PhysRevC.85.044904",
    journal = "Phys. Rev. C",
    volume = "85",
    pages = "044904",
    year = "2012"
}

@article{Frankfurt:2015cwa,
    author = "Frankfurt, L. and Guzey, V. and Strikman, M. and Zhalov, M.",
    title = "{Nuclear shadowing in photoproduction of {\ensuremath{\rho}} mesons in ultraperipheral nucleus collisions at RHIC and the LHC}",
    eprint = "1506.07150",
    archivePrefix = "arXiv",
    primaryClass = "hep-ph",
    doi = "10.1016/j.physletb.2015.11.012",
    journal = "Phys. Lett. B",
    volume = "752",
    pages = "51--58",
    year = "2016"
}

@article{Zha:2021jhf,
    author = "Zha, Wangmei and Tang, Zebo",
    title = "{Discovery of higher-order quantum electrodynamics effect for the vacuum pair production}",
    eprint = "2103.04605",
    archivePrefix = "arXiv",
    primaryClass = "hep-ph",
    doi = "10.1007/JHEP08(2021)083",
    journal = "JHEP",
    volume = "08",
    pages = "083",
    year = "2021"
}

@article{Zha:2023sed,
    author = "Zha, Wangmei and Brandenburg, James Daniel and Xu, Zhangbu",
    title = "{The Breit-Wheeler Process in Relativistic Heavy-Ion Collisions: Creating Matter from Pure Energy}",
    doi = "10.1080/10619127.2023.2230854",
    journal = "Nucl. Phys. News",
    volume = "33",
    number = "3",
    pages = "27--31",
    year = "2023"
}

@article{LHCb:2021hoq,
    author = "Aaij, Roel and others",
    collaboration = "LHCb",
    title = "{$J/\psi$ photoproduction in Pb-Pb peripheral collisions at $\sqrt {s_{NN}}$= 5 TeV}",
    eprint = "2108.02681",
    archivePrefix = "arXiv",
    primaryClass = "hep-ex",
    reportNumber = "LHCb-PAPER-2020-043, CERN-EP-2021-122",
    doi = "10.1103/PhysRevC.105.L032201",
    journal = "Phys. Rev. C",
    volume = "105",
    number = "3",
    pages = "L032201",
    year = "2022"
}

@article{Llope:2014vpd,
    author = "Llope, W. J. and others",
    title = "{The STAR Vertex Position Detector}",
    eprint = "1403.6855",
    archivePrefix = "arXiv",
    primaryClass = "physics.ins-det",
    doi = "10.1016/j.nima.2014.04.080",
    journal = "Nucl. Instrum. Meth. A",
    volume = "759",
    pages = "23--28",
    year = "2014"
}

@article{Selyuzhenkov:2007zi,
    author = "Selyuzhenkov, Ilya and Voloshin, Sergei",
    title = "{Effects of non-uniform acceptance in anisotropic flow measurement}",
    eprint = "0707.4672",
    archivePrefix = "arXiv",
    primaryClass = "nucl-th",
    doi = "10.1103/PhysRevC.77.034904",
    journal = "Phys. Rev. C",
    volume = "77",
    pages = "034904",
    year = "2008"
}

\end{document}